# Describing condensed matter from atomically resolved imaging data: from structure to generative and causal models

Sergei V. Kalinin,[1] Ayana Ghosh,[1,2] Rama Vasudevan,[1] and Maxim Ziatdinov[1,2]

[1]Center for Nanophase Materials Sciences, Oak Ridge National Laboratory, Oak Ridge, TN 37831
[2]Computer and Computational Sciences Division, Oak Ridge National Laboratory, Oak Ridge, TN 37831

The development of high-resolution imaging methods such as electron and scanning probe microscopy and atomic probe tomography have provided a wealth of information on structure and functionalities of solids. The availability of this data in turn necessitates development of approaches to derive quantitative physical information, much like the development of scattering methods in the early XX century which have given one of the most powerful tools in condensed matter physics arsenal. Here, we argue that this transition requires adapting classical macroscopic definitions, that can in turn enable fundamentally new opportunities in understanding physics and chemistry. For example, many macroscopic definitions such as symmetry can be introduced locally only in a Bayesian sense, balancing the prior knowledge of materials' physics and experimental data to yield posterior probability distributions. At the same time, a wealth of local data allows fundamentally new approaches for the description of solids based on construction of statistical and physical generative models, akin to Ginzburg-Landau thermodynamic models. Finally, we note that availability of observational data opens pathways towards exploring causal mechanisms underpinning solid structure and functionality.

# 1. Classical description of solids: symmetry and order parameters

The success of modern condensed matter physics and materials science is deeply rooted in concepts such as symmetries and symmetry-based descriptors, foundational for band theory, quasiparticles, and order parameters.[1-2] These descriptions naturally emerge in systems with discrete translational invariance and intrinsically link to the data available from scattering techniques. Similarly, concepts of macroscopic averaging are ubiquitous across multiple physical disciplines, with average concentrations, electrochemical potentials, etc., being the concepts that form the basis of fields ranging from thermodynamics to materials science and electrochemistry.

Unsurprisingly, this approach is limited for materials that lack long-range translational symmetry including structural, dipole, and spin glasses.[3-4] For some of these systems, the kinetic limitations during the formation process can lead to the formation of out of equilibrium defect populations and frustrated bonded networks. At the same time, geometric frustrations and symmetry mismatch between interacting subsystems can lead to spatially non-uniform ground states that correspond to (one of the) highly degenerate potential energy minima separated by high activation barriers. These materials, as exemplified by spin glasses, ferroelectric relaxors, morphotropic systems, and nanoscale phase separated materials have rapidly become one of the central topics of condensed matter physics, both due to the fundamental physical interest and the unique functional properties they exhibit.[2, 5-6] Strongly correlated electron systems also serve as an active area of research in this community due to presence of charge, spin, orbital, and lattice degrees of freedom, giving rise to competing interactions. A non-exhaustive list of such materials includes conventional superconductors, high-temperature superconductors, magnetic systems (can be frustrated), quantum magnets or quantum Hall systems, and Fermi liquids. While each of these classes of systems set the stage for exciting avenues for research, such solids, in general, have electronic degrees of freedom that produce exotic properties. For example, if we look at doped transition metal oxides with poor screening properties, the interaction energy between valence electrons can overwhelm their kinetic energy, causing a strongly coupled many-body ground state leading to properties such as high-temperature superconductivity and colossal magnetoresistance. Understanding and engineering these interactions are also motivating from the viewpoint of technological applications in efficient energy transmission, electronic devices, optical lattices for quantum communications and computers. Theoretical techniques such as the pseudo particle approach, composite operator method, band structure calculations, projection operator method, dynamical mean field theory, dynamical cluster approximations and spectroscopic methods such as angle-resolved photoemission spectroscopy have been commonly utilized to study such materials.

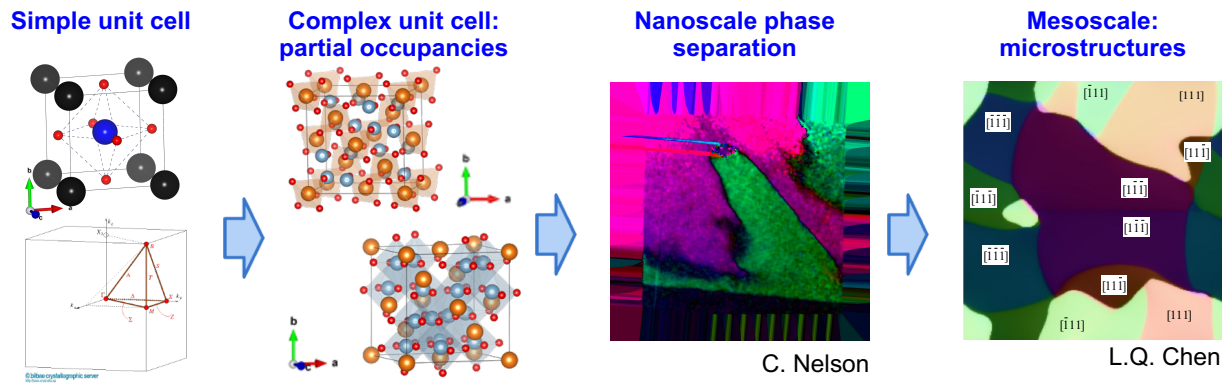

**Figure 1**: General representations of solid systems at different length scales starting from a bulk unit cell, complex unit cell, phase separated nanophases and microstructures.

However, exploring these materials via scattering methods is non-trivial,[7-9] since the average descriptions in the form of correlation functions and structure factors are generally insufficient to reconstruct the exact structure. A prototypical example is a mixture of nanodomains of two higher symmetry phases can appear to show signatures of a low-symmetry phase when the x-ray coherence length is larger than the characteristic domain size. While a variety of stochastic reconstruction techniques have been developed to determine if the structural models satisfying known constraints such as preferred bond lengths and angles and experimentally measured structure factors, such models tend to be non-unique.

## 2. Structural descriptors of solids from local data

The emergence of atomically resolved scanning probe microscopy (SPM) in the last two decades of XX century[10-12] and especially rapid progress in aberration-corrected (Scanning) Transmission Electron Microscopy (STEM)[13-15] and Atomic Probe Tomography (APT) of the last decade has made observation of atomic structures of surfaces and 2D and bulk materials routine. However, availability of the high-resolution imaging data has further necessitated development of methods for description of atomic structures. Traditionally, these have been based on the direct application of the macroscopic concepts such as symmetry, lattice periodicities, etc. For example, the positions and splitting of the Fourier transforms peaks was interpreted as an indication of the symmetry lowering and phase transitions. Techniques such as geometric phase analysis has been used to highlight the violation of periodicity of the lattice and emergence of symmetry-breaking defects.[16-17] However, in these analyses the small details of the symmetry breaking phenomena not associated with translational symmetry are generally ignored.

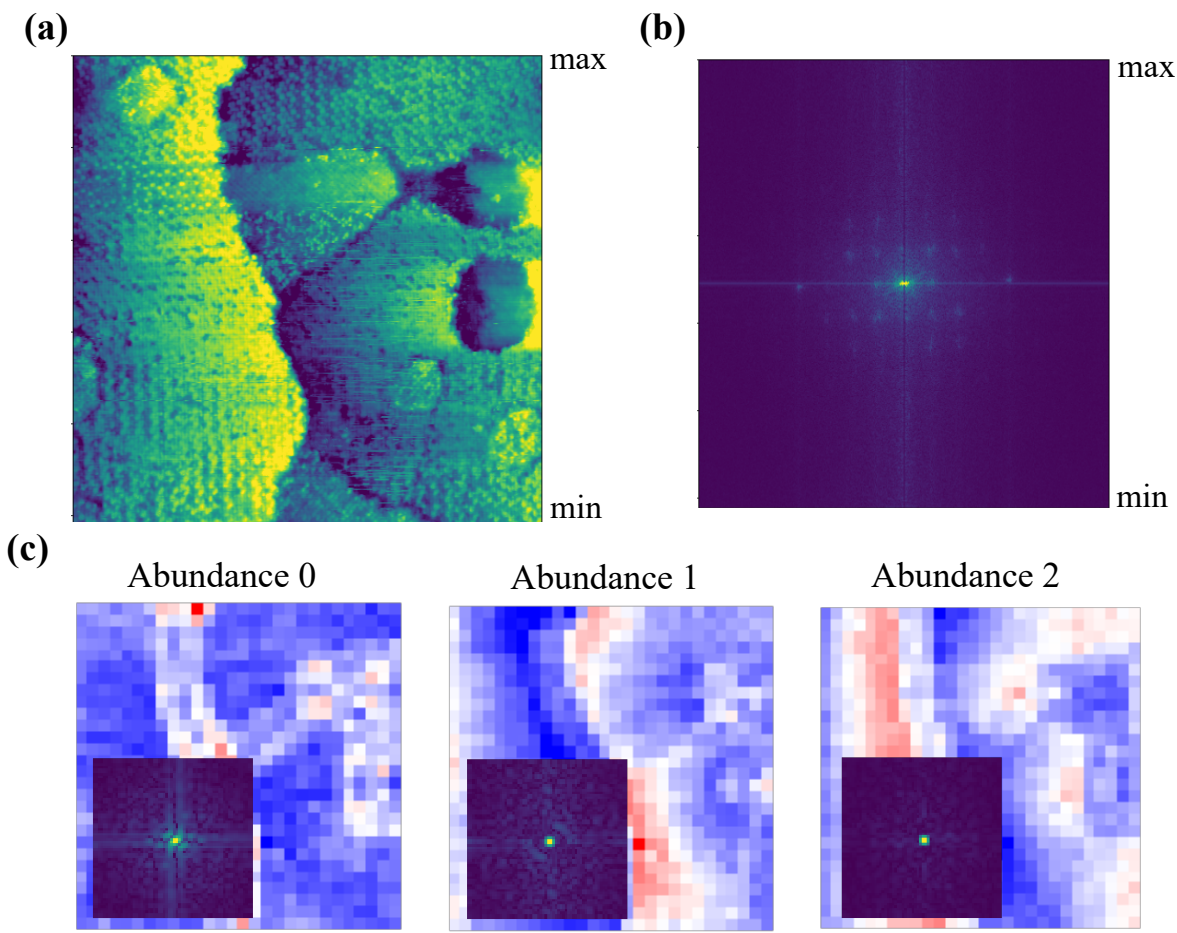

**Figure 2**: Example of discovering features (a-c) from structural data via a combination of reciprocal and real space techniques such as global FFT, sliding FFT and local crystallography. The real space methods are heavily dependent on Gaussian fitting to each atom for the local chemical environment and clustering based on intensities. Techniques such as sliding FFT are dependent on 2D transformations followed by application of dimensionality reduction approaches to unveil local structural information.

This approach can be further extended to the concept of sliding transforms, sometimes referred to as 'texture analysis'.[18-19] Here, the image is sampled by a sliding window of a given size, yielding a collection of (partially overlapping) patches. This approach is somewhat analogous convolutional filters in deep convolutional neural networks and corresponds to the cases when no prior information on the objects of interest is available. The patches in turn can be transformed using 2D fast Fourier Transforms (FFT) or other suitable feature finding methods, and the dimensionality can be further reduced using principal component analysis or other methods. In this manner, the local texture is described via corresponding components, and their variation in the image plane is reflected in maps of weight coefficients ('abundances'). This approach allows for effective segmentation and exploratory data analysis of imaging data, but the connection to physical models is limited by the choice of the transform and the decomposition method. Note that the original choice of the FFT was motivated by several factors, including that the FFT is a direct measure of the periodic structure, that it is linear and therefore linear decomposition methods are also suitably easy to interpret, and lastly the fact that translation converts into the phase of FFT and hence is accounted for. The recent development of the rotational- and shift-invariant variational autoencoders allows for further progress, assuming the shift invariance is sufficient to compensate for sampling.

Alternative approaches for the description of solids have also been proposed, for instance based on the statistical analysis of local atomic neighborhoods. Here, the atomic positions are determined from the observed contrast, and the structure of the nearest neighborhood is analyzed using statistical methods.[20-22] Alternatively, the image patch around the selected atomic features can be analyzed using dimensionality reduction or variational autoencoders.

However, while extremely powerful for exploratory analysis of data, these descriptions do not offer a direct connection to classical macroscopic descriptors. We pose that this can be expected, since macroscopic concepts such as symmetry and symmetry-based descriptors are poorly defined on the local level.[23] For example, observation of a small region of crystalline material is insufficient to unambiguously define the local (plane) symmetry group, since observed atomic positions are accessible only with some error. As another example, it is well known that even in the paraelectric cubic phase of ferroelectric systems such as $BaTiO_3$, there exist small displacements of the central cation, but these cancel on average leading to net-zero polarization.

The classical approach is to obtain a large number of observations and perform a frequency-based analysis. However, the more natural approach for description of solids from local data can be given in the framework of Bayesian models. In this case, the prior knowledge of the material is represented in the form of a prior probability distribution. When no information is available, the prior distribution will be very broad, that is to say, non-informative. When the material identity is known, prior distributions are more narrow (informative priors). Based on observation, the priors are updated to yield the posterior distributions based on Bayes formula.[24] The relationships between the latter and classical physical descriptors (e.g., is the material cubic or tetragonal) can be based on community-accepted criteria. A similar approach can also be applied for the determination of physical parameters from observational data,[25] as discussed below.

One of the simplest applications of Bayesian methods to local structural analysis is a variational autoencoder (VAE). The VAE is a directed latent-variable probabilistic graphical model in which one assumes that complicated real-world observations originate from a small number of the explanatory factors of variation and VAE's aim to recover those (latent) factors (Fig. 3). Here, the optimal results are achieved when the structure of the VAE's latent space takes into account prior domain knowledge. For example, in disordered systems, the same structural blocks may have different orientations in the image plane, which 'confuses' standard decomposition methods such as principal component analysis and non-negative matrix factorization. In the case of VAE, however, one can partition the latent space into the part associated with (random) rotations, and the part associated with the content of the images, and from the latter recover the order parameters in such systems in the unsupervised fashion. Similarly, conditioning the VAE on the already known factors of variation (e.g., partially known discrete classes of structures) allows for more efficient recovery of the unknown ones.

The VAE setup also allows performing certain forms of causal analysis. Generally, the causal links between variables are inferred via a directed acyclic graph (DAG). Since the VAE itself can be represented by a DAG, it is easily extendable to simple causal inference problems. Indeed, Christos *et al.* showed that the latent (hidden) confounders can be recovered given the observed potential confounders, binary treatments, and their outcomes.[26] In addition, a VAE conditioned on an additionally observed variable can be used to discover causal directions in the presence of general non-linear relationships between pairs of passively observed variables via a series of independent tests.[27] Finally, one may learn a DAG in the latent space of the VAE.[28] In this case, the VAE's encoder takes observations as input and generates independent exogenous variables, which are then mapped via a learnable transformation into a causal representation. The latter is fed into the VAE's decoder to reconstruct the observations.

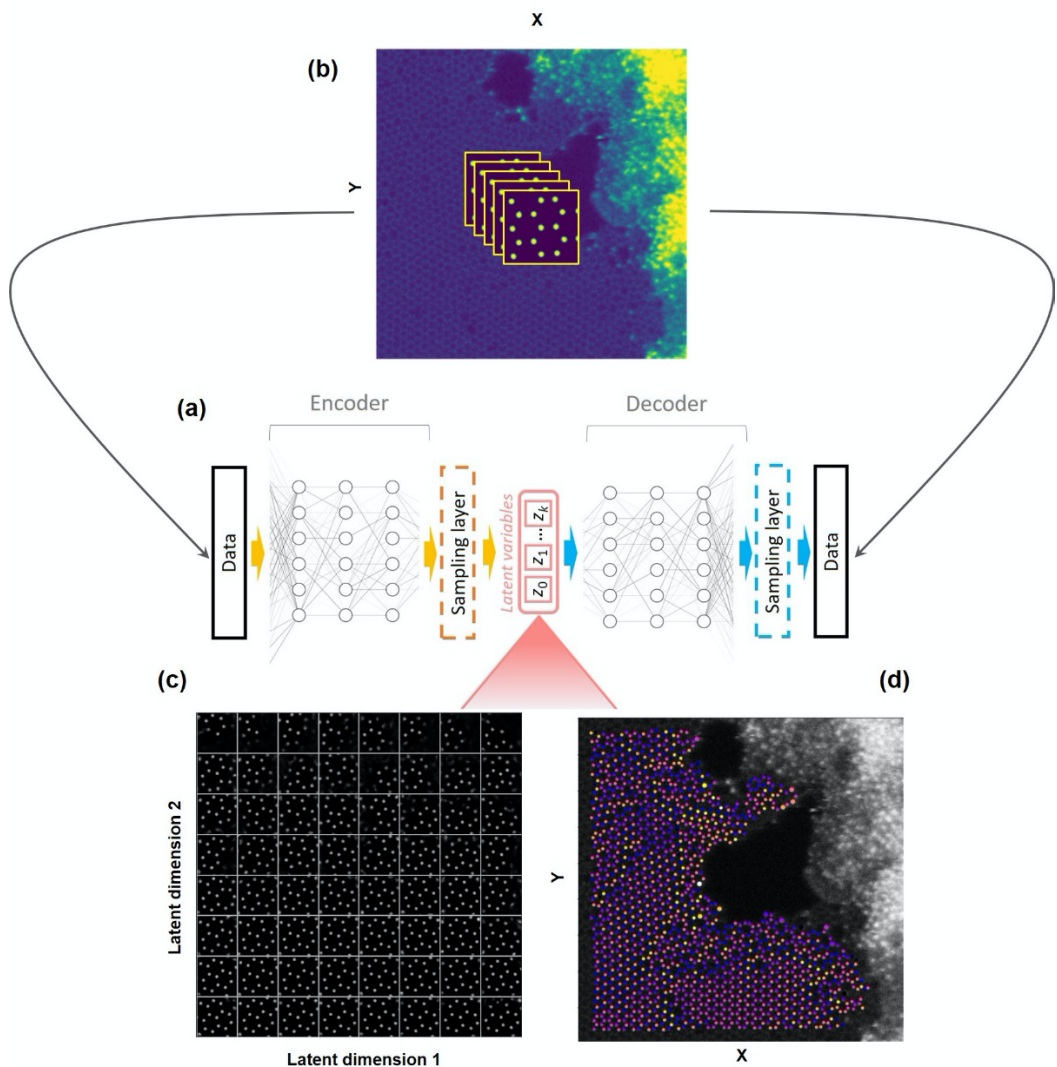

**Figure 3**. Schematic representation of variational autoencoder (VAE) (a). The VAE consists of two Bayesian networks: a generative model (decoder) that tries to recover observations from a latent "code" *z* and an inference model (encoder) that performs a variational inference of the true posterior distribution. An example of a scanning transmission electron microscopy image from graphene monolayer representative image patches used to construct training dataset (b), learned latent manifold (c) and one of the latent variables back-projected to the real space (d) are shown.

## 3. From structure to generative models

It can be argued that one of the goals of microscopic imaging studies is to obtain a statistically significant representation of the structure, meaning that given the observations we can predict how the other (non-observed) regions of the material will appear. For example, the description of solids in terms of the primitive unit cell and ideal discrete translational symmetry allows for exactly this – the knowledge of the single unit cell allows predicting how an arbitrary large volume of material will look like. Naturally, the question emerges whether this approach can be extended to solids lacking long-range periodicity. While approaches such as assembling libraries of structural building blocks are possible,[29] it generally does not allow extrapolation to different regions of parameter space.

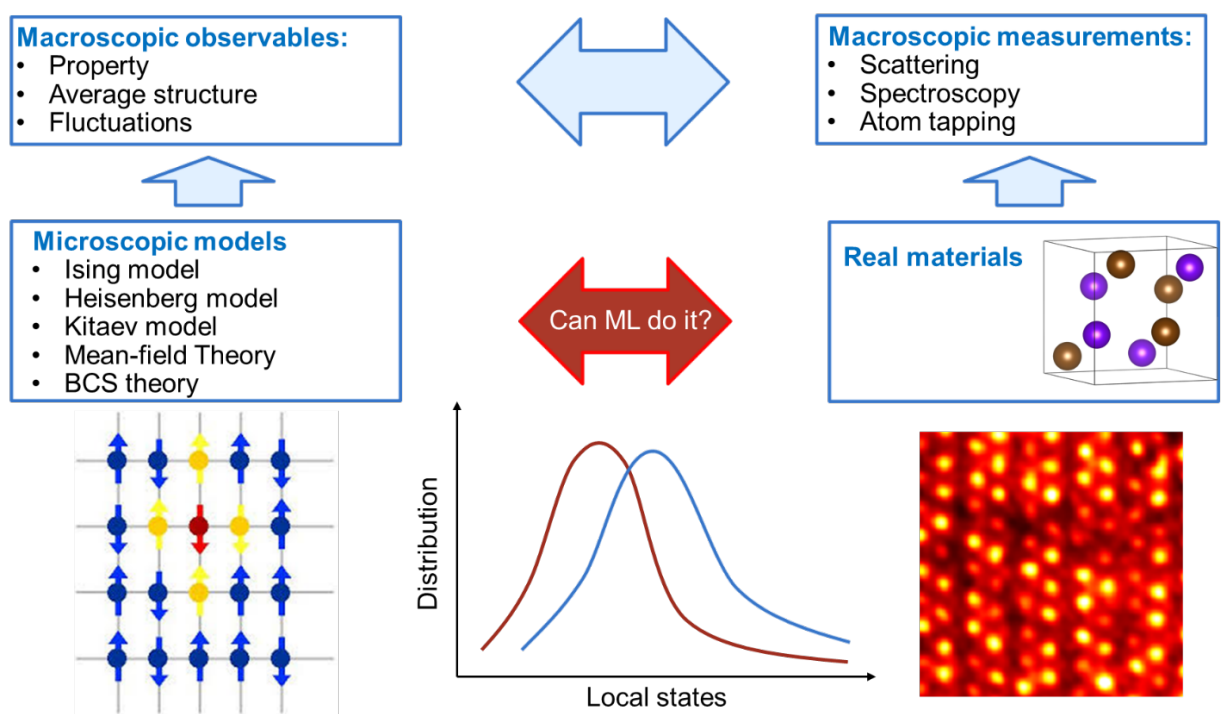

**Figure 4**: This figure represents an overview of the connections between observables and measurements at macroscopic level and how various theoretical models at the microscopic scales are established to understand them and to learn and apply to real materials with respective order parameters controlling functional properties.

It should be noted that this question is not new and has been considered: the traditional language of crystallography is insufficient to explain even the structure of water ice; these systems display correlated disorder where systems do not possess long-range periodicity but do possess local ordering.[23] In such situations the space group formalism fails, but correlations in disorder can still be determined through diffraction techniques via analysis of diffuse scattering, e.g., using pair distribution functions (PDF). Interestingly, these correlations are found across many different systems, pointing to the possibility of a unifying framework to understand correlated disorder regardless of the interactions driving them in each individual case. For example, symmetry mismatch between the lattice and the interactions can give rise to specific form of correlated disorder that is true regardless of material type and has the same signatures in diffuse scattering. Nonetheless, it is still a major challenge to describe disordered matter, and major progress still relies on expensive calculations of different possible structures from which PDFs can be computed and compared with experiment.

An alternative approach to descriptions of solids is one based on a generative model that gives rise to the equivalent stochastic atomic, dipole, or spin microstructure. In other words, rather than describing the structure of a solid, we can aim to reconstruct the generative model that gives rise to observed microstructures and will correctly describe the system in the statistical sense. Note that this is one step above the 'average' structure that will yield the correct PDF; however, with it lies the challenge of needing to collect appropriate statistics on local atomic configurations.

The generative adversarial network (GAN) approach to generative models has some similarity to this concept but is based on correlative properties. With GANs, the main idea is to model the underlying generative distribution by a dueling network architecture, wherein the generative model (typically a neural network) attempts to produce samples that mimic those samples in a training set and seeks to defeat a critic. This critic (also usually a neural network) aims to correctly classify those data originating from the generative model, and those ones belonging to the training set. The Nash Equilibrium for this competitive game is that the generator eventually becomes able to mimic the samples in the training dataset, thus not allowing the critic anything more than a random chance at determining which distribution the test sample was drawn from. Mathematically this is written as the problem of minimizing the Jennson-Shannon (JS) divergence between the two distributions: the one being modeled by the generator, and the true distribution estimated from sampling of the training samples.

One recent advance to GANs is the use of the Wasserstein GAN, which reformulates the loss function to incorporate a Wasserstein metric as opposed to the traditional JS metric and this appears to improve training. One interesting aspects of GANs is they have a neat parallel between methods to derive physical models. Indeed, one method to know whether a physical model adequately describes an experiment is if the sample drawn from the physical generative model are distinguishable from those drawn from the experiment. This should be calculated via statistical distance optimization: to model observables in a physical system, we can optimize the parameters of a generative model to minimize the statistical distance between the derived probability distributions of the relevant observations. This method accounts for uncertainty in sampling, as well as respecting thermodynamics. As such, learning the underlying generative model that can be sampled, and then directly compared with local observations, provides a route towards understanding disordered materials without needing to rely on the difficulties presented by scattering data.

Hence, learning the generative model from atomically resolved data, incorporating prior knowledge, and yielding corresponding uncertainties as posterior parameter distributions is a clear opportunity. Similarly, the size of the descriptor can be an issue - for example the latter, will diverge in the vicinity of the phase transitions.

However, this goal can be recast more broadly, as understanding the rules that govern the emergence of the structure, rather than structure itself. As a special case of describing a solid structure is enumerating the deterministic rules by which the structure is formed. For ideal crystals these rules are trivial – just continuous tiling of the identical blocks. The example of non-trivial rules will then be quasicrystals. Notably, the areas such as cellular automata serve as an example of the corresponding mathematical framework. The obvious limitation here is that while if the rules controlling the formation of a solid are a known prediction, it is generally straightforward (forward model), but the determination of the formation rules from the observations of the structure is considerably more complex problem.

Here, we pose that the target of the emerging status can be the reconstruction of the Hamiltonian of the system, or in other words a generative physical model that gives rise to the structure, rather than structure itself. On the mesoscopic level, this can be accomplished by the direct matching between mesoscopic Ginzburg-Landau type models and local imaging data. This approach has been used to explore the nature of the boundary and gradient terms in ferroics and vacancy in ordered systems,[30] as well as the flexoelectric constants.[31] In these, the mesoscopic model acts as a regularizer providing a smooth envelope over atomic data. These methods can be extended to incorporate the prior knowledge via Bayesian methods in a straightforward manner.[25]

In the mesoscopic world, phase-field modeling is a common approach that works with the material's compositions profile as a field in space allowing for modeling polycrystalline microstructure reconstruction, solidification, kinetics of coarsening, grain boundary migration and grain growth, by using numerical methods to integrate partial differential equations (PDEs) within appropriate boundary conditions (BCs). This differs from the conventional idea of phase separations that can easily be modeled by free energy functions as a function of temperature. For physical phenomena such as spinodal decompositions at interfaces, chemical compositions may vary continuously between different regions. The order parameter is reevaluated and extended to phases and thus becomes a function of space along with temperature. Interfaces between phases have state variables that are intermediate between adjacent regions, domains. In the Ginzburg-Landau equation which is often considered as the starting point for any phase field model, free energy density is expressed in terms of an order parameter and its gradient. This, when integrated over volume leads to Cahn-Hilliard equation used to model spinodal decomposition. It is important to mention that this order parameter may very well be a non-conserved quantity like magnetic spin alignment, which may change in one region without being compensated in the other. Tensor fields such as strain fields, vector fields such as magnetic fields and other field quantities along with non-linear energetic relationships existing between these field quantities are generally included to represent a complete model for a complex microstructure. Overall, solving these PDEs numerically which may vary for different thermodynamic processes can turn out to be computationally challenging. Generative models to unlock mesoscopic features and produce meaningful models will thereby be beneficial. These models can serve as guiding tools to account for a large set of empirical variables, drawing statistical inferences to guide the reconstruction or other processes, sampling over a broad space to produce a good representative model of a material of interest.

The phase-field model itself contains numerous parameters, and the ability to include or exclude many different terms that contribute to the free energy such as the elastic, electric, gradient of these. For appropriate modeling, it is necessary to estimate these parameters from experiment, which can be difficult. An example is where direct imaging of vortices in PTO/STO enabled determination of the flexoelectric coefficient via computer-vision approaches by Li et al.[31] More generally, understanding the appropriate model parameters becomes a Bayesian model selection problem, but which again requires large computation over the available model space. Alternative strategies include fully differentiable systems to optimize parameters in a similar way to back propagation in neural networks. While maintaining data quality and size itself could be challenging for engineering related applications, introducing physics laws into the deep learning frameworks can assist with such problems. However, inclusion of the appropriate PDEs into deep learning frameworks is also not straightforward, due to the nonlinearity and nonlocality of the PDEs. This is also tied to the conventional forward approach of solving the PDEs to obtain a representation of material or physical phenomena. Recent studies[32-36] have shown that there is a possibility to construct a top-down or inverse method to investigate integral relations of complex systems directly from macroscopically observed fields via PDE constrained optimization in a Bayesian setting by using images and patterns found in them. Here, a smart sampling approach such as Markov Chain Monte Carlo (MCMC) can be used to sample over a broad parameter space to explore and exploit the datapoints in cases of both conserved and non-conserved order parameters for estimation of free energy landscapes. Taking the inverse of PDEs that are capable of explaining the macroscopic pattern-formation dynamics, would readily connect to various macroscopic properties from the microscopic images with associated uncertainties. This inversion approach has excellent applications in formalizing electro-chemical to thermal processes and in evaluating processes that go beyond the standard descriptions of the phase-field models.

We pose that this approach is natural for atomically resolved systems. For example, binary solid solutions can be fully described via the corresponding Ising-like Hamiltonian that gives rise to statistically-similar (in a sense of some metric, e.g. Kullback-Leibler divergence of distributions) microstructures. This description is compact (akin to Kolmogorov complexity of a Mandelbrot set), and generalizable to non-observed concentrations and temperatures. It has been shown that direct observation of the mesoscopic degrees of freedom can be directly compared to a lattice model via statistical distance minimization.[37-39] We have also explored the veracity of the statistical distance minimization and associated uncertainty quantification for the paradigmatic Ising model and demonstrated that in the presence of weak bond-disorder the exchange integral can be determined well above associated bulk phase transitions.

This approach can be considerably extended. For the cases with strong inductive biases (e.g., symmetry of the order parameter), we can explore the use of the strongly-structured priors in the probabilistic models of observed phenomena. These models, as exemplified e.g., by latent variable Gaussian process or equivalent deep generative models, assume the presence of latent variables in the system and (unknown) data generation processes that give rise to experimental observations. This type of framework naturally comports to the postulate of the existence of an order parameter, and an imaging process affected by the instrumental factors. The second approach is the introduction of global thermodynamic energy based non-translation invariant kernels in the composition space. In this case, the uncertainty in the optimization process comprises both that of the experiment and global model. Finally, the VAE approach described in Section 2 can be used not only to recover a latent representation of the system but also as a generative model. Here, assuming that each latent variable is associated with an independent physical factor of variation, one can generate structures corresponding to their different combinations.

## 4. From static to dynamic systems

In many cases, dynamic system behavior is experimentally observable, including mesoscopic and atomic-scale observations of phase transitions, ferroelectric domain dynamics, and direct atomic motion induced by chemical, thermal, and bias stimuli. In certain cases, the observed dynamics are induced by observational process, e.g., electron beam-stimulated atomic dynamics in STEM or ferroelectric domain switching in Piezoresponse Force Microscopy. Hence, the natural question is whether these observations can yield additional information on materials.

Here, the fundamental factors determining the level of additional insight in materials functionality is determined by sampling in the time domain and the possibility for intervention. For example, observations of dynamic beam induced processes in solids allow collecting information on the metastable atomic configurations, if these can be formed via beam irradiation. However, interpretation of this statistical information in terms of relevant physics is complicated.

From the modeling perspective, ab-initio molecular dynamics (AIMD) can provide tremendous amounts of information on structural and physical behavior, but it is limited by the length and time scales. Classical molecular dynamics (MD) simulations can be thought of as an alternative to address the challenge of length and time scale although a compromise needs to be made in terms of achieving quantum mechanical level accuracy as determined by interatomic potentials. ML-learned potentials combined with a Bayesian framework to learn from both observational, simulation data (trajectories from every step of MD simulation), that modify the parameters in the potentials to reconstruct structure-property diagrams have also paved their way in this community.[40-44] Furthermore, a counterfactual approach to represent a molecule without much initial understanding of the feature space and mapping it back to real structures can also prove to be beneficial in terms of performing dynamic interventions of rather 'unknown' systems in experimental or simulation environments.[45]

Comparatively, observation of the (differentiable) trajectories allows much deeper insight into generative physics of the system. Over last several years, the work by Battaglia[36] have demonstrated the potential for discovery of the equations of motion from the observation of celestial motion. Similarly, Kutz[46-48] and others have shown discovery of ordinary and partial differential equations from imaging data.

Starting from describing physio-chemical processes to understanding spatiotemporal evolution of complex systems to approximate underlying microscopic observations, these are often dependent on constructing models using macroscopic observables and solving ordinary different equations (ODEs) and PDEs. However, this can turn out to be mathematically rigorous and time-consuming as briefly mentioned earlier in the previous section. It is possible to first consider the coarse-grain fields of a system (such as diffusion maps) to extract relevant macroscopic features in the form of coarse derivatives that map existing relations of these features with time to then predict PDEs with generative models.

A special case is the one where controlled intervention is possible. A trivial case is PFM, where the domain formation is induced by the tip and the relationship between the size of the formed domain and bias can be established. A less trivial example is the electron beam atomic dynamics, which in special cases can be (partially) controlled.[49-50]

## 5. Generative models is not enough: causality in physical processes

A basic underpinning of physical sciences is their causal nature, with an intuitively understood cause and effect relationship. However, deep understanding and implication of causal phenomena are generally not analyzed. For example, in thermodynamic equilibrium the causal links are degenerate; however, any dynamic process has a cause and effect. Certain physical models incorporate the causal link implicitly, e.g., Kramers-Kronig relationships. Here we hypothesize that the knowledge of the generative mechanisms is generally insufficient to understand the physics of complex materials systems with multiple competing interactions. The generative physical model can yield a statistically significant representation of material structure including all relevant subsystems under the conditions of thermodynamic equilibrium, or ergodic systems.

However, many material systems including ferroelectric relaxors or morphotropic materials are non-ergodic, as shown in Figure 5. In this case is it is insufficient to provide a generative model; rather, cause and effect mechanisms that lead to the emergence of materials microstructure should be analyzed. For example, can we argue that in morphotropic and relaxor ferroelectrics that (frozen) cation distributions pin the polarization order parameter field, or polarization field instabilities drive cation (and oxygen vacancy) distributions? Similarly, can we establish whether the average electron concentration or local chemistry drives distortion patterns in doped $MX_2$? Some of these questions can be answered by macroscopic experiment (e.g., electron concentration effects and chemical environment can be separated by electrostatic gating), but in some cases direct experiments are expensive or impossible.

In certain cases, these questions can be answered by macroscopic experiments. For example, electron concentration effects and chemical environments can be separated by electrostatic gating in field effect devices. Similarly, ferroelectric instability can be tuned independently (of composition) by pressure. However, in many cases direct experiments are expensive or impossible, or are associated with additional effects. For example, electrostatic gating to change electron concentration often couples to chemical changes in the material.

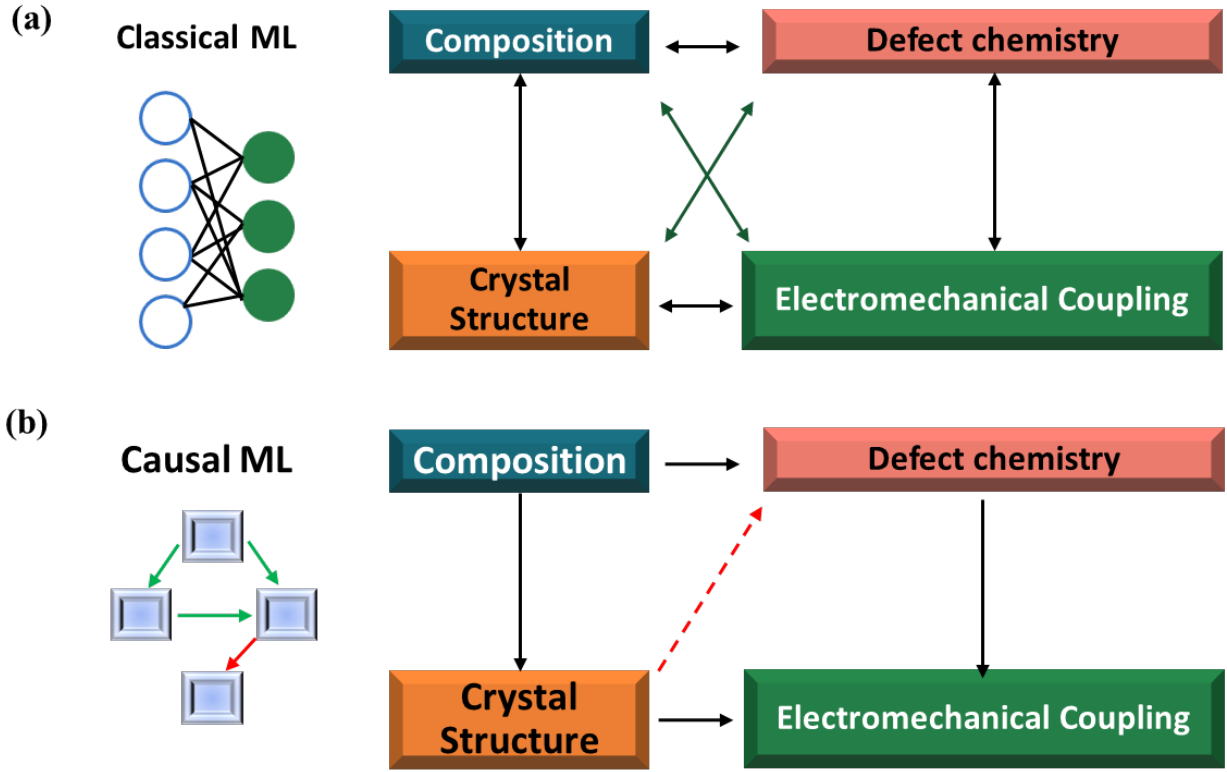

**Figure 5.** Causal and correlative machine learning for ferroelectric relaxors via global composition, local composition, local polarization, and electromechanical response. In (a) classical ML, established are the correlative relationships between the variables based on domain specific expectations, and the causal links are either postulated or derived from regression coefficients, often leading to incorrect interpretations. (b) In causal ML, the task is the discovery of the unknown or partially known causal graph and establishing the functional links between the nodes. Note that in active causal learning the concept is close to that of regular physical experiment, with the controlled modification of the system and analysis of the relevant outcomes.

We argue that observational data under certain conditions can be used to establish these causal relationships, thus advancing our understanding of these materials. [51] The systematic frameworks for studies of the causal mechanisms in complex materials using graph model formalisms include those developed by Pearl,[52-56] Scholkopf,[57-60] Lopes-Paz,[61-62] Bottou,[63] and others.

However, the true power of these probabilistic methods is that arbitrary physics-based DAGs can be in principle defined in the latent space of the VAEs. Indeed, the classical VAE structure described earlier assumes the absence of the causal structure among the factors of variation, that is, the observations are assumed to be generated by independent latent factors. Hence, it may not be adequate in situations where there can be a causal relationship between the latent factors of interest. As mentioned earlier, the causal links are generally inferred via a directed acyclic graph or DAG, Fig. 6. In the case of VAE, the standard latent variables $z_i$ can be interpreted as noise driving a structural casual model[64] and the causal relationships can be modeled via $s_i =$ $f_i(\boldsymbol{pa_i}, z_i)$ where $s_i \ldots s_n$ are associated with the vertices of DAG, and $f_i$ is a deterministic function (e.g., a neural network) that depends on the $s_i$'s parents in the graph, $\boldsymbol{pa_i}$, and the encoded latent variables.[28] Even though a full recovery of the causal structure may not always be possible, a representation that partially exposes the relevant causal links could be useful for downstream tasks of interest.

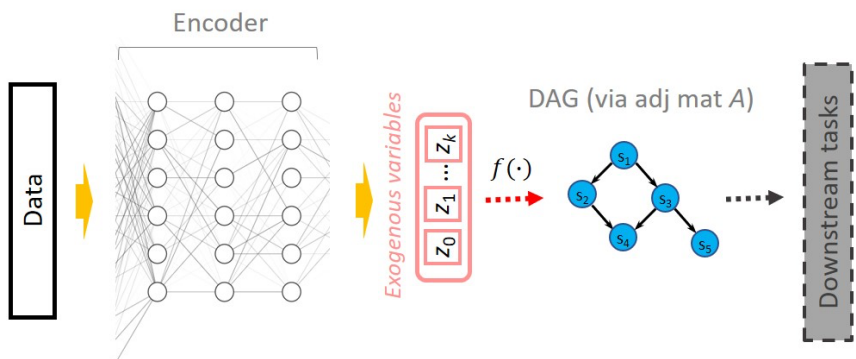

**Figure 6**. Schematic of the causal learning with VAE. For brevity, only the encoder part is shown (i.e., we assume an already trained VAE model) and no sampling layers are depicted. Here, the VAE latent variables are the unexplained noise variables that are mapped into a causal "layer" via a learnable transformation. The fully or partially exposed causal structure can be used for the downstream tasks of interest.

We note that currently the existing causal inference models support only directed acyclic causal graphs, whereas many physical phenomena are associated with cyclic graph models. However, the causal ML community have started to explore such models[65] and, given the very rapid progress in the field, we expect these capabilities to be available in the forthcoming year. Together, physical and causal models offer a pathway for knowledge-based materials design, including interventional and counterfactual questions, as well as guiding experiments.

## 6. Summary

The impact of the adoption of these concepts can be expected to be threefold. First, it will set a common language for the structural analysis of imaging data, obviating multiple existing controversies. Secondly, adoption of generative models as descriptors for solid microstructure will establish a pathway towards the utilization of multiple existing data, and also allow much closer connection between the experiment and theory realms via machine learning tools. Finally, compact physics based and causal models by their nature allow for generalization, extrapolation, and answering counterfactual questions – what is likely to happen if we substitute a specific atom, apply pressure, etc. – the ultimate measure of physical knowledge.

This effort is based upon work supported by the U.S. Department of Energy (DOE), Office of Science, Basic Energy Sciences (BES), Materials Sciences and Engineering Division (S.V.K., R.K.V.), U.S. DOE, Office of Science, Office of Basic Energy Sciences Data, Artificial Intelligence and Machine Learning at DOE Scientific User Facilities (A.G.). and was performed and partially supported (M.Z.) at the Oak Ridge National Laboratory's Center for Nanophase Materials Sciences (CNMS), a U.S. Department of Energy, Office of Science User Facility.